%% file: main.tex
\documentclass[preprint,10pt]{elsarticle}
\usepackage[margin=1in]{geometry}
\usepackage{graphicx} 
\usepackage{bm}
\usepackage{amsfonts,amsmath}
\usepackage{amsthm}
\usepackage{thmtools,thm-restate}

\usepackage{mathtools}
\usepackage{xcolor}
\usepackage{natbib}
\usepackage{todonotes}
\usepackage[normalem]{ulem}
\numberwithin{equation}{section}

\usepackage{float}
\input{definitions.tex}

\journal{Elsevier}
\usepackage{ifthen}
\newcommand{\ShowComments}{false}
\newcommand{\forus}[1]{\ifthenelse{\equal{\ShowComments}{true}}{{\footnotesize \color{blue!90!black}~[[#1]]}}{}}

\begin{document}

\title{Determining the acceleration field of a rigid body using three accelerometers and one gyroscope, with applications in mild traumatic brain injury}
\author[1]{Yang Wan}
\author[1,2]{Benjamin E.~Grossman-Ponemon}
\author[1]{Haneesh Kesari\corref{cor1}}
\ead{haneesh\_kesari@brown.edu}
\address[1]{School of Engineering, Brown University, Providence, RI 02912, USA}
\address[2]{New address: Department of Physics and Engineering, John Carroll University, University Heights, OH 44118, USA}
\date{}

\cortext[cor1]{Corresponding author }

\begin{abstract}
Mild traumatic brain injury (mTBI) often results from violent head motion or impact. Most prevention strategies explicitly or implicitly rely on motion- or deformation-based injury criteria, both of which require accurate measurements of head motion.
We present an algorithm for reconstructing the full acceleration field of a rigid body from measurements obtained by three tri-axial accelerometers and one tri-axial gyroscope. Unlike traditional gyroscope-based methods, which require numerically differentiating noisy angular velocity data, or gyroscope-free methods, which may impose restrictive sensor placement or involve nonlinear optimization, the proposed algorithm recovers angular acceleration and translational acceleration by solving a set of linear equations derived from rigid body kinematics.
In the proposed method, the only constraint on sensor placement is that the accelerometers must be non-collinear.
We validated the algorithm in controlled soccer heading experiments, demonstrating accurate prediction of accelerations at unsensed locations across trials. The proposed algorithm provides a robust, flexible, and efficient tool for reconstructing rigid body motion, with direct applications in contact sports, robotics, and biomechanical injury prediction.
\end{abstract}

\maketitle




\section{Introduction}
\label{sec:IntroPreamble}

Mild traumatic brain injury (mTBI) is a highly prevalent neurological condition.
In the US and Europe, $190-225$ patients die following a traumatic brain injury (TBI) event, with tens of thousands more developing chronic neurodegenerative conditions and long-term complications as a result \cite{TBIdata, Maas2022}.
Mild traumatic brain injuries are typically caused by violent head motions, which may result from blunt impacts to the head sustained during contact sports, military activities, motor vehicle accidents, falls, or blast exposures.
Affected individuals often experience a spectrum of cognitive and physical symptoms, including dizziness, headaches, memory disturbances, and impaired recognition, which can persist well beyond the acute phase of injury \cite{Marshall257,Dikmen2017,YEATES2010}.

In mTBI, the primary injury results from the violent motion of the head, leading to the deformation of brain tissue, stretching and tearing of blood vessels, neurons, glia, and axons. This initial trauma subsequently triggers the onset of secondary injury processes, including metabolic dysfunction, inflammation, and neurodegeration \cite{masel2010traumatic,xiong2013,CERNAK2005,GAETZ2004,BRAMLETT2007125}.
Given the high incidence of mTBI and its potential for long-term neurological consequences, there is an urgent need for effective prevention and mitigation strategies in mechanically induced traumatic events.

Most mTBI prevention strategies explicitly or implicitly rely on a "brain injury criterion" for their effective systhesis, implementation, and evaluation. These criteria take as input various descriptors of head motion—such as linear acceleration, angular velocity, angular acceleration or mechanical responses of brain tissue, including strains and strain rates.
Based on these biomechanical inputs, the brain injury criterion provides an estimate of the likelihood that a given head motion or deformation following a mechanical trauma event will result in mTBI.

Examples of motion-based brain injury criteria include
the Gadd Severity Index (SI) and the Head Injury Criterion (HIC) \cite{gadd1966use,chou1974analytical}, both of which take the head's translational acceleration as their input. In contrast, the Brain Injury Criterion (BrIC) \cite{takhounts2013development} and GAMBIT \cite{newman1986generalized} consider the head's rotation motion's potential to cause the injury. The input to GAMBIT is the angular acceleration time series and the input to Bric is the head's angular velocity time series.
These criteria are empirically derived and relatively straightforward to apply, as they involve evaluating simple algebraic expressions.

More recently, brain deformation-based injury criteria have been developed. These criteria are based on the hypothesis that the risk of brain injury is correlated with brain tissue strains and strain rates \cite{hajiaghamemar2021multi, bar2016strain,RafaelTBImodel,wan2023mechanics}.
To support this approach, a wide range of computational mechanics-based head models have been created to simulate the brain’s deformation response to head motion during traumatic events.
These models vary in complexity—from high-resolution models that incorporate detailed brain geometry and account for the anisotropic mechanical properties of the brain (e.g., \cite{madhukar2019finite, Carlsen2021}) to more simplified models with lower spatial resolution and reduced anatomical detail (e.g., \cite{Massouros2014, wan2023brain}). Most of these models rely on finite element methods to solve the governing equations of brain deformation and stress in response to mechanical loading.
To enable faster estimation of brain deformation, recent advances have introduced machine learning-based methods that directly predict brain strain from head kinematics data \cite{upadhyay2022data, zhan2021rapid, wu2022real}.




Irrespective of whether a head-motion-based or brain-deformation-based brain injury criterion is
used, both rely on the availability of a robust algorithm for accurately reconstructing head motion.
Head-motion-based criteria use motion descriptors derived from measurable data as input. On the other hand, the brain-deformation-based criteria require the time history of head motion to serve as the loading conditions in computational head models in order to
calculate the resulting strains and strain rates.
There are existing algorithms capable of fully determining the motion of the head by using measurements from inertial sensors including accelerometers and gyroscopes. These sensors can be conveniently mounted on personnel's head via wearable sensor system, enabling the measurement of kinematics over expansive spaces.
These sensor systems, including instrumented custom mouthguards, earpieces, skin patches, and
headbands \cite{Jeneel2025,tripathi2025laboratory,Buice2018,abrams2024}.

In parallel with advancements in sensor technology, numerous sensor data processing algorithms have been developed to estimate the acceleration field of the head based on sensor measurements.
In the context of general rigid body motion, the acceleration field in terms of the body frame (which is a set of vectors that are attached to the rigid body, and hence move with it), is calculated as (\cite[3.4]{wan2022determining})
\begin{equation} \busf{A}\ag{\tau}\ag{\usf{X}}=\busf{W}\ag{\tau}\busf{W}\ag{\tau}\usf{X}+\busf{W}'\ag{\tau}\usf{X}+\usf{q}\ag{\tau},
    \label{equ:psuedoacce}
\end{equation}
where $\tau\in \mathbb{R}$ denotes a non-dimensional time instant; $\usf{X}\in \mathbb{R}^3$ denotes a rigid body material particle; $\busf{A}\ag{\tau}\ag{\usf{X}}\in \mathbb{R}^3$ is the material particle' acceleration in body frame; $\busf{W}\ag{\tau}$ is a time dependent square matrix of real numbers that denotes the angular velocity of the rigid body in body frame; $\busf{W}'\ag{\tau}$ is the derivative of $\busf{W}\ag{\tau}$;
$\usf{q}\ag{\tau}\in \mathbb{R}^3$ is a time dependent vector. Finally the acceleration in laboratory frame can be calculated once we know how the body frame rotates w.r.t. the laboratory frame.



Algorithms for estimating a rigid body's acceleration field from sensor measurements are typically categorized into two types: gyroscope-based and gyroscope-free approaches.
In gyroscope-based algorithm, the body-frame angular velocity matrix $\busf{W}\ag{\tau}$ is directly obtained from gyroscope measurement.
However, these methods commonly compute the body-frame angular acceleration $\busf{W}'\ag{\tau}$ by numerically differentiating the measured angular velocity time series (e.g., \cite{Camarillo2013}).
It is well known that numerical differentiation tends to significantly amplify measurement noise, which can adversely affect the accuracy of the resulting acceleration estimates \cite{ovaska1998noise, alonso2005noise}.

Gyroscope-free algorithms estimate the acceleration field by measuring the acceleration at multiple points on the rigid body using accelerometers  \cite{rahaman2020accelerometer,wan2022determining,cardou2008estimating,cardou2009linear}. Assuming the spatial positions of these accelerometers are known, a system of linear equations can be formed based on Equation \eqref{equ:psuedoacce}.
The unknown terms, $\busf{W}\ag{\tau}\busf{W}\ag{\tau}$, $\busf{W}'\ag{\tau}$, and $\usf{q}\ag{\tau}$ can then be solved for linearly, provided that twelve independent equations are available, for example, from twelve uniaxial accelerometers, six dual-axial accelerometers, or four triaxial accelerometers.
Rahaman~\textit{et
al.}~\cite{rahaman2020accelerometer} and Wan~\textit{et
al.}~\cite{wan2022determining} developed an algorithm for determining the acceleration field using data only from four tri-axial accelerometers.

Although gyroscope-free methods avoid the noise amplification associated with numerical differentiation in gyroscope-based approaches, they come with their own limitations.
Some algorithms impose
constraints on the locations and orientations of the accelerometers, requiring specific configurations
to ensure accurate motion predictions.
For instance, the method of Padgaonkar~\textit{et
al.}~\cite{padgaonkar1975measurement} require that accelerometers are oriented
perpendicular to one another and located along each accelerometer’s measurement axes. Zou and Angeles \cite{zou2018algorithm} require each accelerometers to be located at the centroid of each face of the polyhedron.
In contrast, the methods proposed in \cite{wan2022determining,rahaman2020accelerometer} allow more flexible sensor placement, as long as all accelerometers do not lie on the same plane.
Xiong \cite{Xiaobo2025} and Naunheim~\textit{et al.}~\cite{naunheim2003linear} impose comparatively relaxed requirements on the placement of accelerometers, with the primary constraint being that the sensors must not be collinear.
However, their methods estimate the parameters associated with the acceleration field through nonlinear fitting procedures, which require solving a set of nonlinear equations or performing numerical minimization—potentially increasing computational complexity and sensitivity to initialization or noise.

In this work, we proposed a novel algorithm to determine the acceleration field of a rigid body. In our algorithm, we use measurement data from three triaxial accelerometer and one triaxial gyroscope, hence we refer to the proposed algorithm as the A3G1-algorithm.
The proposed method imposes minimal constraints on sensors placement and orientation: the only requirement is that the three accelerometers must not lie along the same line.
In the A3G1-algorithm, the body-frame angular velocity $\busf{W}\ag{\tau}$ is directly obtained from the gyroscope measurement. The body-frame angular acceleration $\busf{W}'\ag{\tau}$ and the vector $\usf{q}\ag{\tau}$ are then computed by solving a set of linear equations derived from \eqref{algo:acce}, using the acceleration measurements from the three accelerometers.
As a result, the A3G1-algorithm avoids the drawbacks of numerical differentiation and nonlinear optimization, offering a robust and computationally efficient solution.

The paper is structured as follows. The mathematical preliminaries necessary for detailing our algorithm are described in \S\ref{sec:prelim}. We develop the proposed algorithm, A3G1-algorithm in \S\ref{sec:algorithm}.
We presented the implementation procedure of the A3G1-algorithmin in \S\ref{sec:algo} and illustrate the algorithm with some examples.
In \S\ref{sec:demo} we showcase an experimental demonstration of our algorithm to determine the acceleration of the player's head in a soccer header.
Concluding remarks are provided in \S\ref{sec:ConcludingRemarks}.

\section{Preliminary mathematics and kinematics of rigid body motion}
\label{sec:prelim}

In this section we briefly recapitulate the mathematics and kinematics of rigid body motion~\cite{rahaman2020accelerometer,wan2022determining,wan2023brain}. 
For simplicity, we use a non-dimensional, matrix formulation, rather than the physical or dimensional formulation of \cite{rahaman2020accelerometer,wan2022determining,wan2023brain}. The interested reader may view \cite{rahaman2020accelerometer,wan2022determining,wan2023brain} for the more complete formulation.

\subsection{Definitions and notation}
\label{sec:notation}

\begin{figure}[h]
    \centering
        \includegraphics[width=\textwidth]{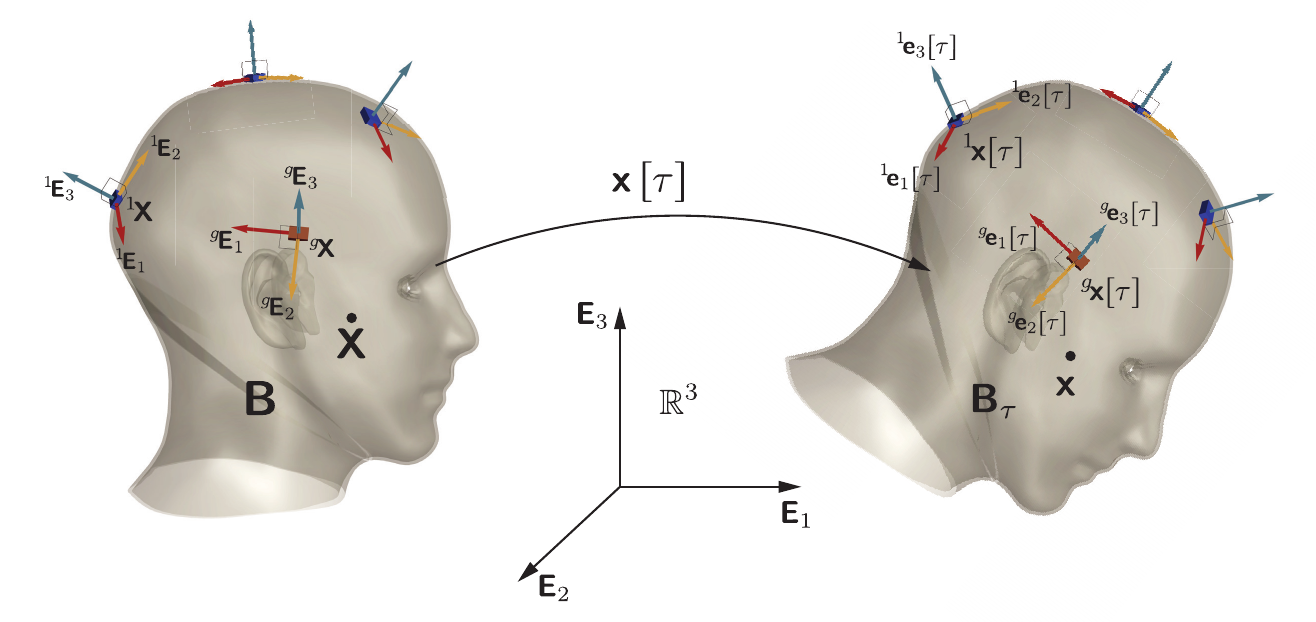}
    \caption{Mathematical quantities used in the description of motion and schematic of the locations and orientations of four tri-axial accelerometers (left) and their motion (right).
  }
    \label{fig:AOillu}
\end{figure}

Let $\usf{B} \subset \mathbb{R}^3$ be a rigid body, which we call the reference configuration, or body frame.
We call an element $\usf{X} \in \usf{B}$ the reference point (see Fig.~\ref{fig:AOillu}).
We model the
rigid body's motion using the one-parameter family of maps $\usf{x}\ag{\tau}\ag{\cdot}:\usf{B}\rightarrow \mathbb{R}^3$. We call $\usf{x}\ag{\tau}$ the deformation map and $\usf{x}\ag{\tau}\ag{\usf{X}}$ the reference point $\usf{X}$'s current position at the time instance $\tau$ (see Fig.~\ref{fig:AOillu}). When there is no confusion, we abbreviate $\usf{x}\ag{\tau}\ag{\usf{X}}$ as $\usf{x}$.
The set $\usf{B}_{\tau}=\set{\usf{x}\ag{\tau}\ag{\usf{X}}\in\mathbb{R}^3}{\usf{X}\in \usf{B}}$ is called the current body.

Let $\pr{\usf{E}_{i}}_{i\in\mathcal{I}}$, where $\mathcal{I}:=\pr{1,2,3}$, be an orthonormal set of basis vectors for $\mathbb{R}^3$. By orthonormal we mean that the inner product between $\usf{E}_i$ and $\usf{E}_j$, where $i, j\in\mathcal{I}$, equals $\delta_{ij}$,
the Kronecker delta symbol, which equals unity if $i=j$ and zero otherwise.
We denote the space of all $m \times n$ real matrices, where $m, n\in \mathbb{N}$, $\mathcal{M}_{m \times n}(\mathbb{R})$;
here $\mathbb{N}$ and $\mathbb{R}$ denote the set of natural numbers
and the space of real numbers, respectively. Thus, $\usf{X}\in\mathcal{M}_{3\times1}(\mathbb{R})$, or $\mathbb{R}^3$.
We access the $i$-$j^{\rm th}$ component of $\usf{A}\in \mathcal{M}_{m \times n}(\mathbb{R})$, where $i\in \pr{1,\ldots,m}$, $j\in \pr{1,\ldots,n}$, as $\Absf_{\cdot i\cdot j}$.

\paragraph{Combination of matrices}
We will abuse notions to denote matrices which are comprised of other matrices. For example, let  $\Absf\in \mathcal{M}_{m\times n}(\mathbb{R})$, $\Hbsf \in \mathcal{M}_{m\times p}(\mathbb{R})$, and $\Mbsf\in \mathcal{M}_{q\times n}(\mathbb{R})$. Then we write
$$
\begin{aligned}
&\begin{pmatrix}
\Absf & \Hbsf
\end{pmatrix} =
&\begin{pmatrix}
\Absf_{\cdot 1\cdot 1} & \cdots & \Absf_{\cdot 1\cdot n} & \Hbsf_{\cdot 1\cdot 1} & \cdots & \Hbsf_{\cdot 1\cdot p}  \\
\vdots & \ddots & \vdots & \vdots & \ddots & \vdots  \\
\Absf_{\cdot m\cdot 1} & \cdots & \Absf_{\cdot m\cdot n} & \Hbsf_{\cdot m\cdot 1} & \cdots & \Hbsf_{\cdot m\cdot p}
\end{pmatrix} \in \Mc_{m \times (n+p)}(\mathbb{R}),
\end{aligned}
$$
and
$$
\begin{aligned}
&\begin{pmatrix}
\Absf \\ \Mbsf
\end{pmatrix} =
&\begin{pmatrix}
\Absf_{\cdot 1\cdot 1} & \cdots & \Absf_{\cdot 1\cdot n}   \\
\vdots & \ddots & \vdots  \\
\Absf_{\cdot m\cdot 1} & \cdots & \Absf_{\cdot m\cdot n}   \\
\Mbsf_{\cdot 1\cdot 1} & \cdots & \Mbsf_{\cdot 1\cdot n}   \\
\vdots & \ddots & \vdots  \\
\Mbsf_{\cdot q\cdot 1} & \cdots & \Mbsf_{\cdot q\cdot n}
\end{pmatrix} \in \Mc_{\pr{m+q} \times n}(\mathbb{R}).
\end{aligned}
$$

\subsection{Rigid body motion}
\label{sec:rbd}

For the case of rigid body motion $\usf{x}\ag{\tau}$ takes the form
\begin{equation}
\usf{x}\ag{\tau}\ag{\usf{X}}=\usf{Q}\ag{\tau}\usf{X}+\usf{c}\ag{\tau},
\label{equ:deformmap}
\end{equation}
where $\usf{c}:\mathbb{R}\rightarrow \mathbb{R}^3$ is twice (Fr\'{e}chet) differentiable and $\usf{Q}\ag{\tau} \in \mathcal{M}_{3\times3}(\mathbb{R}): \mathbb{R}^3 \rightarrow \mathbb{R}^3$ is a proper rotation and belongs to the special orthonormal group $ SO(3)\subset \mathcal{M}_{3\times 3}\pr{\mathbb{R}}$, satisfying the equations
\begin{subequations}
\begin{align}
\usf{Q}^{\sf T}\ag{\tau}\,\usf{Q}\ag{\tau}&=\usf{I}_{3\times3},\label{equ:QQ:1}
\intertext{and}
\usf{Q}\ag{\tau}\,\usf{Q}^{\sf T}\ag{\tau}&=\usf{I}_{3\times3},\label{equ:QQ:2}
\end{align}
\label{eq:QQ}
\end{subequations}
where $\usf{Q}^{\sf T}\ag{\tau}$ is the transpose of $\usf{Q}\ag{\tau}$, i.e., $\usf{Q}^{\sf T}\ag{\tau}=\pr{\usf{Q}\ag{\tau}}^{\sf T}$ and $\usf{I}_{3\times3}=\pr{\delta_{ij}}_{i,j\in\mathcal{I}} \in \mathcal{M}_{3 \times 3}(\mathbb{R})$.

\paragraph{\textbf{Velocities}}
Taking the derivative of the function $\usf{x}\ag{\cdot}\ag{\usf{X}}$ w.r.t. $\tau$, we get the velocity of $\usf{X}$. It follows from \eqref{equ:deformmap} that
\begin{equation}
\usf{V}\ag{\tau}\ag{\usf{X}}:= \usf{Q}'\ag{\tau}\usf{X}+\usf{c}'\ag{\tau}.
\label{equ:velocity}
\end{equation}
Using \eqref{equ:deformmap} and \eqref{equ:velocity}, it can be shown that the velocity at the time instance $\tau$ of the point occupying the current position $\usf{x}\in \mathbb{R}^3$ is $\usf{W}\ag{\tau}\pr{\usf{x}-\usf{c}\ag{\tau}}+\usf{c}'\ag{\tau}$, where $\usf{W}\ag{\tau}\in \mathcal{M}_{3 \times 3}(\mathbb{R})$ is defined by
\begin{equation}
\usf{W}\ag{\tau}=\usf{Q}'\ag{\tau}\usf{Q}^{\sf T}\ag{\tau}.
\end{equation}
Here the matrix $\usf{W}\ag{\tau}$ belongs to the space of real skew-symmetric matrices $\mathfrak{so}(3)\subset \mathcal{M}_{3\times 3}\pr{\mathbb{R}}$.
We can associate with $\usf{W}\ag{\tau}$ the matrix $\usf{w}\ag{\tau}\in \mathbb{R}^3$ that is defined such that
\begin{equation}
   \usf{W}\ag{\tau}\usf{x}= \usf{w}\ag{\tau}\times \usf{x},
\end{equation}
for all $\usf{x} \in \mathbb{R}^3$, where the symbol ``$\times$" denotes the cross product operator. The matrix $\usf{w}\ag{\tau}$ is interpreted as angular velocity of the rigid body at time instance $\tau$.
The relation between $\usf{W}\ag{\tau}$ and $\usf{w}\ag{\tau}$ can also be expressed using the map $\star\ag{\cdot}:\mathfrak{so}(3) \rightarrow \mathbb{R}^3$ defined by the equation
\begin{equation}
\star\ag{\cdot}=\pr{\pr{\cdot}_{32},\pr{\cdot}_{13},\pr{\cdot}_{21}}.
\label{eq:star}
\end{equation}
It can be shown that, $\usf{w}\ag{\tau}=\star\ag{\usf{W}\ag{\tau}}$.
The inverse of $\star\ag{\cdot}$ is the map $*\ag{\cdot}:\mathbb{R}^3\to\mathfrak{so}(3)$
defined by the equation
\begin{equation}
*\ag{a_1,a_2,a_3}=
\begin{pmatrix}
0 & -a_3 & a_2\\
a_3 & 0 & -a_1\\
-a_2 & a_1 & 0
\end{pmatrix}.
\label{eq:astar}
\end{equation}
It can be shown that for $\usf{A}\in \mathfrak{so}(3)$ and $\usf{x} \in \mathbb{R}^3$,
\begin{equation}
    \usf{A}\usf{x}=\pr{*\ag{\usf{x}}}^{\sf T}\star\ag{\usf{A}}.
    \label{equ:hodge1}
\end{equation}

\paragraph{\textbf{Accelerations}}
Taking the second derivatives of the function $\usf{x}\ag{\cdot}\ag{\usf{X}}$ w.r.t. $\tau$, we get the acceleration of $\usf{X}$, which can be shown that
\begin{equation}
\usf{A}\ag{\tau}\ag{\usf{X}}:= \usf{Q}''\ag{\tau}\usf{X}+\usf{c}''\ag{\tau}.
\label{eq:acce}
\end{equation}

Finally we acquire the ``Pseudo-acceleration field'' (which can be interpreted as the acceleration components in body frame), which is defined as
\begin{equation}
\busf{A}\ag{\tau}\ag{\usf{X}}:= \usf{Q}^{\sf T}\ag{\tau}~\usf{A}\ag{\tau}\ag{\usf{X}}.
\label{eq:pseudoacce}
\end{equation}
It then follows \eqref{eq:acce} and \eqref{eq:pseudoacce} that
\begin{equation}
\busf{A}\ag{\tau}\ag{\usf{X}}:= \usf{P}\ag{\tau}\usf{X}+\usf{q}\ag{\tau},
\label{eq:pseudoacce1}
\end{equation}
where $\usf{P}\ag{\tau}:=\usf{Q}^{\sf T}\ag{\tau}\usf{Q}''\ag{\tau}$ and $\usf{q}\ag{\tau}:=\usf{Q}^{\sf T}\ag{\tau}\usf{c}''\ag{\tau}$.

As shown in \cite[1.2 and 1.3]{wan2022determining}, the matrix $\usf{P}\ag{\tau}$ has the form,
\begin{equation}
\usf{P}\ag{\tau}= \busf{W}\ag{\tau}\busf{W}\ag{\tau}+\busf{W}'\ag{\tau},
\label{eq:p}
\end{equation}
\begin{equation}
\busf{W}\ag{\tau}= \star\ag{\busf{w}\ag{\tau}},
\label{eq:Wbar}
\end{equation}
where $\busf{w}\ag{\tau}\in \mathbb{R}^3$ is the pseudo-angular velocity of the rigid body at the time instance $\tau$ and it is related to the angular velocity $\usf{w}\ag{\tau}$ through the equation (\cite[2.59]{rahaman2020accelerometer}),
\begin{equation}
\usf{w}\ag{\tau}= \usf{Q}\ag{\tau}\busf{w}\ag{\tau}.
\label{eq:wbar}
\end{equation}
In our proposed algorithm, $\busf{w}\ag{\tau}$ can be determined from the gyroscope measurement.

Thus, the Pseudo-acceleration field $\busf{A}\ag{\tau}$ is taken to be fully determined once $\usf{P}\ag{\tau}$ and $\usf{q}\ag{\tau}$ have been computed.

\section{Proposed algorithm}
\label{sec:algorithm}
We propose an algorithm for determining the acceleration field of the rigid body $\Bbsf$ using measurements from three tri-axial accelerometers and one tri-axial gyroscope. We consider the accelerometers and gyroscope are rigidly attached to $\Bbsf$. We denote the reference points of  $\Bbsf$, to which these three accelerometers are attached, as  $\pr{\lsc{\ell}\usf{X}}_{\ell\in\mathcal{I}}$, respectively, and the gyroscope as $\lsc{g}\usf{X}$.

Each accelerometer measures the components of its acceleration in three mutually orthogonal directions. We refer to those directions as the accelerometer's measurement axis. In the reference configuration, we denote the measurement axes of accelerometer $\lsc{\ell}\usf{X}$ as $\pr{\lsc{\ell}\usf{E}_i}_{i\in\mathcal{I}}$. As the body $\Bbsf$ moves, the attached accelerometers move with it, and the measurement axes will change with time. In the current configuration at time instance $\tau$, accelerometer $\lsc{\ell}\usf{X}$ has current position $\lsc{\ell}\usf{x}\ag{\tau} = \usf{x}\ag{\tau}\ag{\lsc{\ell}\usf{X}}$ and measurement axes $\pr{\lsc{\ell}\usf{e}_i\ag{\tau}}_{i\in \mathcal{I}}$, where $\lsc{\ell}\usf{e}_i\ag{\tau} = \Qbsf\ag{\tau}\lsc{\ell}\usf{E}_i$ for each $i\in \mathcal{I}$.

The gyroscope works in the same way as accelerometer does, instead it measures the components of its angular velocity in three mutually orthogonal directions.
We denote its measurement axes in reference configuration, measurement axes in current configuration, and current position as $\pr{\lsc{g}\usf{E}_i}_{i\in\mathcal{I}}$, $\pr{\lsc{g}\usf{e}_i[\tau]}_{i\in\mathcal{I}}$, and $\lsc{g}\usf{x}[\tau]$, respectively. The reference points $\pr{\lsc{\ell}\usf{X}}_{\ell\in\mathcal{I}}$, $\lsc{g}\usf{X}$, and the directions $\pr{\lsc{\ell}\usf{E}_i}_{i,\ell\in\mathcal{I}}$, $\pr{\lsc{g}\usf{E}_i}_{i\in\mathcal{I}}$ are known from the arrangement and orientation of the accelerometers and gyrocope at the experiment's starting.

Let $\lsc{\ell}\alpha_{i}\ag{\tau}$, $i\in\mathcal{I}$, be the measurement reported by accelerometer $\lsc{\ell}\usf{X}$ for the (non-dimensional) component of its acceleration in the $\lsc{\ell}\usf{e}_i\ag{\tau}$ direction at the time instance $\tau$.
Then it can be shown that the pseudo-acceleration at the reference point $\lsc{\ell}\usf{X}$ at the time instance $\tau$, $\lsc{\ell}\busf{A}(\tau):=\busf{A}\ag{\tau}\ag{\lsc{\ell}\usf{X}}$ is
\begin{equation}
\lsc{\ell}\busf{A}(\tau)= \pr{\usf{E}_i^{\sf T}\sum_{j=1}^{3}\lsc{\ell}\alpha_{j}\ag{\tau}\lsc{\ell}\usf{E}_j}_{i \in \mathcal{I}}.
\label{eq:accei}
\end{equation}

Let $\lsc{g}\omega_{i}\ag{\tau}$, $i\in\mathcal{I}$, be the measurement reported by the gyroscope $\lsc{g}\usf{X}$ for the (non-dimensional) component of its angular velocity in the $\lsc{g}\usf{e}_i\ag{\tau}$ direction at the time instance $\tau$.
As for a rigid body, every point has the same angular velocity.
Then it can be shown that the pseudo-angular velocity of the rigid body at the time instance $\tau$ is
\begin{equation}
\busf{w}\ag{\tau}= \pr{\usf{E}_i^{\sf T}\sum_{j=1}^{3}\lsc{g}\omega_{j}\ag{\tau}~\lsc{g}\usf{E}_j}_{i \in \mathcal{I}}.
\label{eq:angvel}
\end{equation}

It follows from \eqref{eq:pseudoacce1} and \eqref{eq:p} that
\begin{equation}
    \lsc{\ell}\busf{A}(\tau)=\busf{W}\ag{\tau}\busf{W}\ag{\tau}\lsc{\ell}\usf{X}+\busf{W}'\ag{\tau}\lsc{\ell}\usf{X}+\usf{q}\ag{\tau}.
    \label{equ:psuedoacceell1}
\end{equation}
In the right hand side of \eqref{equ:psuedoacceell1}, write $\busf{W}'\ag{\tau}\lsc{\ell\,}\usf{X}$ as
$\pr{*\ag{\lsc{\ell}\usf{X}}}^{\sf T}\star\ag{\busf{W}'\ag{\tau}}$ following \eqref{equ:hodge1} and in the resulting equation, substitute $\star\ag{\busf{W}'\ag{\tau}}$ with $\busf{w}'\ag{\tau}$, we can get
\begin{equation}
    \lsc{\ell}\busf{A}\ag{\tau}=\busf{W}\ag{\tau}\busf{W}\ag{\tau}\lsc{\ell}\usf{X}+\pr{*\ag{\lsc{\ell}\usf{X}}}^{\sf T}\busf{w}'\ag{\tau}+\usf{q}\ag{\tau}.
    \label{equ:psuedoacceell2}
\end{equation}

Let $\usf{M}:\mathbb{R}\rightarrow \Mc_{9 \times 1}(\mathbb{R})$,
\begin{equation}
\usf{M}\ag{\tau}=\begin{pmatrix} \lsc{1}\busf{A}(\tau)-\busf{W}\ag{\tau}\busf{W}\ag{\tau}\lsc{1}\usf{X}\\
\lsc{2}\busf{A}(\tau)-\busf{W}\ag{\tau}\busf{W}\ag{\tau}\lsc{2}\usf{X}\\
\lsc{3}\busf{A}(\tau)-\busf{W}\ag{\tau}\busf{W}\ag{\tau}\lsc{3}\usf{X}
\end{pmatrix},
\label{eq:M}
\end{equation}
$\usf{D
}\in \Mc_{9 \times 6}(\mathbb{R})$,
\begin{equation}
    \usf{D
}=\begin{pmatrix}\pr{*\ag{\lsc{1}\usf{X}}}^{\sf T}  &\usf{I}_{3\times3}\\
\pr{*\ag{\lsc{2}\usf{X}}}^{\sf T} & \usf{I}_{3\times3}\\
\pr{*\ag{\lsc{3}\usf{X}}}^{\sf T}  &\usf{I}_{3\times3}
\end{pmatrix},
\label{eq:D}
\end{equation}
and
$\usf{S}:\mathbb{R}\rightarrow \Mc_{6 \times 1}(\mathbb{R})$,
\begin{equation}
 \usf{S}\ag{\tau}=\begin{pmatrix}
\busf{w}'\ag{\tau}\\
\usf{q}\ag{\tau}
\end{pmatrix}.
\end{equation}
It can be shown using \eqref{equ:psuedoacceell2} that
\begin{equation}
\label{equ:matrix-form}
\usf{D
}~\usf{S}\ag{\tau}=\usf{M}\ag{\tau}.
\end{equation}

Here we try to obtain $\usf{S}\ag{\tau}$ by solving the equation \eqref{equ:matrix-form}. As \eqref{equ:matrix-form} is overdetermined, there may not exist an exact solution.
An alternate strategy for computing $\usf{S}\ag{\tau}$ from \eqref{equ:matrix-form} is based on finding a best approximant for $\usf{S}\ag{\tau}$ in $\Mc_{6 \times 1}(\mathbb{R})$.
let this approximant be $\hat{\usf{S}}\ag{\tau} \in \Mc_{6 \times 1}(\mathbb{R})$ and
\begin{equation}  \hat{\usf{S}}\ag{\tau}=\argmin_{\usf{Y} \in \Mc_{6 \times 1}(\mathbb{R})}\norm{\usf{D
}\usf{Y}-\usf{M}\ag{\tau}}_2,
\end{equation}
where $\norm{\cdot}_2$ denotes the standard Euclidean norm.
Following \cite[\S5.3]{golub2013matrix},
assuming $\usf{D}$ has full column rank, i.e., $\usf{D}^{\sf T}\usf{D}$ is invertible,
we get
\begin{equation}
    \hat{\usf{S}}\ag{\tau}=\pr{\usf{D}^{\sf T}\usf{D}}^{-1}\usf{D}^{\sf T}\usf{M}\ag{\tau}.
    \label{equ:solution}
\end{equation}

Knowing the approximants $\busf{w}'\ag{\tau}$ and $\usf{q}\ag{\tau}$ from \eqref{equ:solution}, and  $\busf{w}\ag{\tau}$ from \eqref{eq:angvel}, $\busf{A}\ag{\tau}\ag{\usf{X}}$ is completely determined by \eqref{eq:pseudoacce1}, \eqref{eq:p}, and \eqref{eq:Wbar}.

\subsection{Implementation procedure}
\label{sec:algo}
Following \S\ref{sec:algorithm}, we summarize the procedure to calculate the body-frame acceleration field in Algorithm~\ref{algo:acce}.

\RestyleAlgo{ruled}
\SetKwComment{Comment}{/* }{ */}
\begin{algorithm}[h]
\caption{Computing the body-frame acceleration, $\overline{\usf{A}}\ag{\tau}\ag{\usf{X}}$, of the point $\usf{X}$ at time instance $\tau$}
\label{algo:acce}
\vspace{3pt}
\textbf{Input}:
\
\begin{enumerate}
\item The position of the point of interest $\usf{X}$\;
\item Positions and directions of each accelerometer, $\lsc{\ell}\usf{X}$ and $\pr{\lsc{\ell}\usf{E}_i}_{i \in \mathcal{I}}$, $\ell\in\mathcal{I}$;
\item Directions of the gyroscope, $\pr{\lsc{g}\usf{E}_i}_{i \in \mathcal{I}}$\;
\item Acceleration measurements from three accelerometers at a discrete sequence of time instances, $\pr{\lsc{\ell}\alpha_i\ag{\tau}}_{i \in \mathcal{I}}$, $\ell\in\mathcal{I}$, and angular velocity measurements from the gyroscope at a discrete sequence of time instances, $\pr{\lsc{g}\omega_i\ag{\tau}}_{i \in \mathcal{I}}$\;
\end{enumerate}
\textbf{Output}: the body-frame acceleration, $\overline{\usf{A}}\ag{\tau}\ag{\usf{X}}$, of the point $\usf{X}$ at time instance $\tau$\;
\vspace{3pt}
\textbf{Procedure}:
\begin{enumerate}
    \item Compute the body-frame angular velocity at time instance $\tau$, $\busf{w}\ag{\tau}$,  using \eqref{eq:angvel}\;
    \item Compute $\busf{W}\ag{\tau}$,  using \eqref{eq:Wbar}\;
    \item Compute the body-frame acceleration at each accelerometer $\lsc{\ell}\usf{X}$ at time instance $\tau$, $\lsc{\ell}\busf{A}\ag{\tau}$, $\ell \in \mathcal{I}$ using \eqref{eq:accei}\;
    \item Compute the matrices $\usf{M}\ag{\tau}$ using \eqref{eq:M} and $\usf{D}$ using \eqref{eq:D}\;
    \item Compute the matrix $\hat{\usf{S}}\ag{\tau}$ using \eqref{equ:solution}\;
    \item  Following \eqref{equ:psuedoacceell2}, calculate $\overline{\usf{A}}\ag{\tau}\ag{\usf{X}}$ as  $\busf{W}\ag{\tau}\busf{W}\ag{\tau}\usf{X}+\pr{*\ag{\usf{X}}}^{\sf T}\busf{w}'\ag{\tau}+\usf{q}\ag{\tau}$\;
\end{enumerate}
\Return{$\overline{\usf{A}}\ag{\tau}\ag{\usf{X}}$.}
\end{algorithm}

\section{Demonstration of A3G1 algorithm}
\label{sec:demo}
In this section, we demonstrate the usage of A3G1 algorithm to determine the acceleration field of a player's head in soccer heading exercises. We give an overview of the soccer heading test in \S\ref{subsec:test}, apply the A3G1-algorithm in \S\ref{sec:implementation}, and show the results in \S\ref{subsec:results}.

\subsection{Soccer header test overview}
\label{subsec:test}
\begin{figure}[h]
    \centering
        \includegraphics[width=1\textwidth]{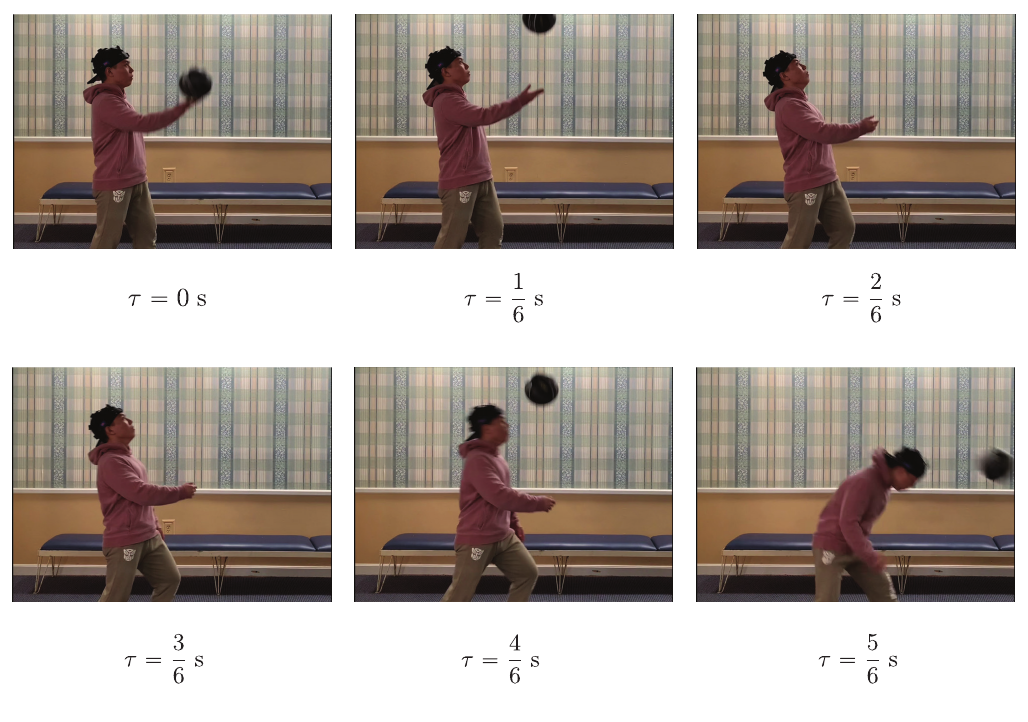}
    \caption{Soccer header test.
  }
    \label{fig:testov}
\end{figure}

The soccer heading exercise is illustrated in Fig.~\ref{fig:testov}, where an individual throws a soccer ball and subsequently heads it.
The individual was the first author of this manuscript. His head circumference, height, and weight are 59~{\rm cm}, 170~{\rm cm}, and 70~{\rm kg}, respectively
In the test, we collected the acceleration and angular velocity data using a wearable, head-mounted sensor system, shown in Fig.~\ref{fig:testov1}(a).
The wearable sensor system consists of a rigid sensor holder with mounting points for four inertial measurement units (IMUs), shown as blue cuboid in Fig.~\ref{fig:testov1}(a).
A more abstract representation of our sensor system is shown in Fig.~\ref{fig:testov1}(b).
Here we use the IMU developed by Vicon, which contains a tri-axial accelerometer and a tri-axial gyroscope.
We denote the accelerometers of each IMU as $A_1, \ldots, A_4$, and gyroscopes of each IMU as $G_1, \ldots, G_4$ (marked, e.g., in Fig.~\ref{fig:testov1}(a)).
The sensor holder was 3D printed using fused deposition molding (FDM, material: polylactic acid (PLA), 3D printer: Original Prusa i3 MK3, Prusa Research, the Czech Republic).
The parameters in its design were used to compute the reference positions and orientations of IMUs. These values for our test are given in the caption of Fig.~\ref{fig:testov1}.


We conducted the soccer heading test five times. The sensor measurements from the representative trial, Test \#3, are presented in Fig.\ref{fig:testov1} (e--f).
Acceleration measurements from accelerometers $A_1$, $A_3$, and $A_4$ are shown in Fig.~\ref{fig:testov1} (c)--(e), respectively.
Angular velocity measurements from gyroscope $G_1$ are shown in Fig.~\ref{fig:testov1}(f).
Each subfigure shows graphs of three (discrete) functions. For example, subfigure (d) shows graphs of $\lsc{3}\alpha_i\ag{\cdot}$, $i=1,2,3$.

\subsection{Applying the A3G1-algorithm}
\label{sec:implementation}
\begin{figure}[H]
    \centering
        \includegraphics[width=1\textwidth]{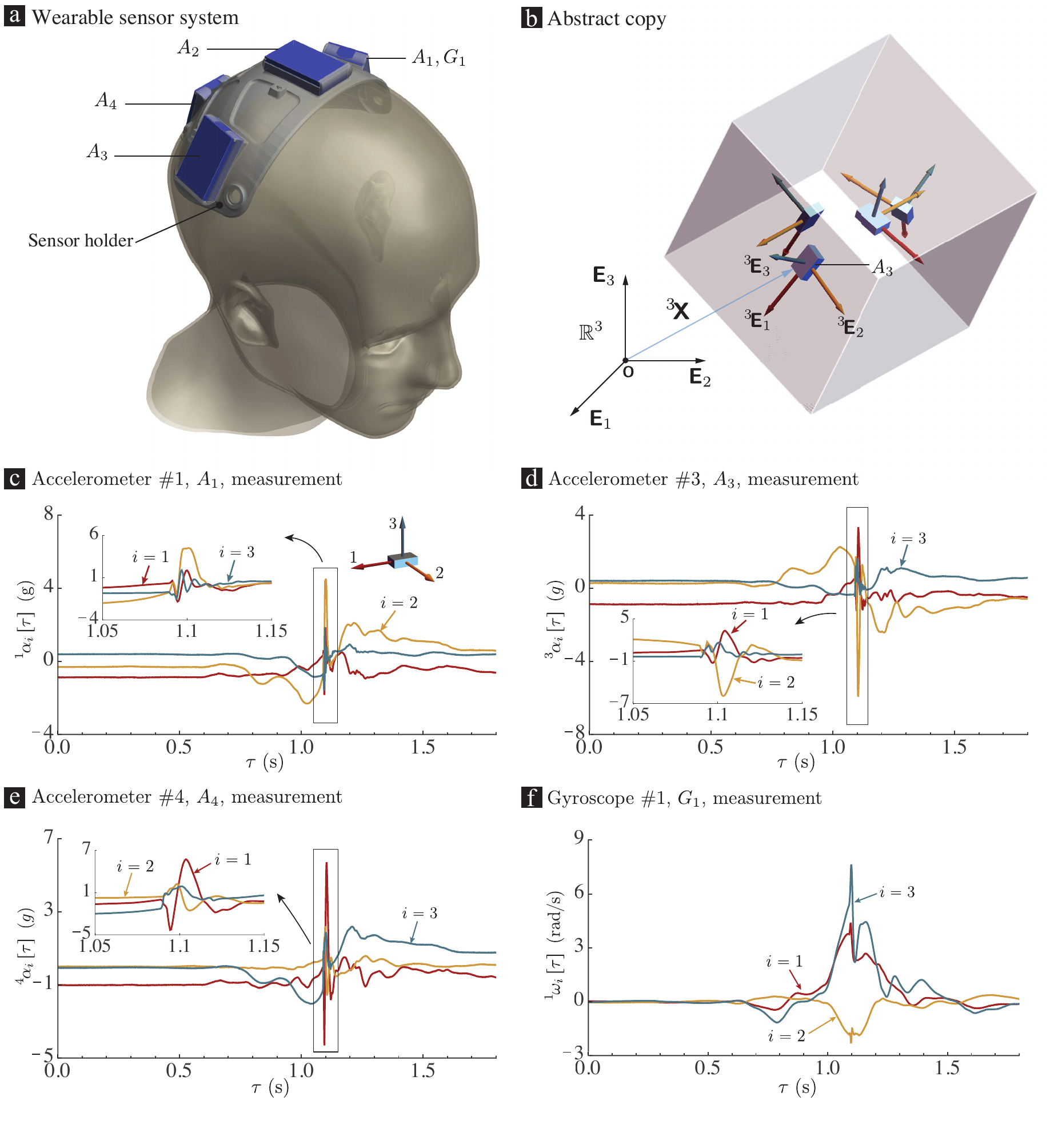}
    \caption{Soccer header experiment setup and sensor measurements. (a) shows a sketch of the wearable sensor system used in the header experiment. (b) shows a more abstract version of the wearable sensor system in the reference configuration. In our experiment, we choose the basis vectors of $\mathbb{R}^3$ as $\usf{E}_1=\pr{1,0,0}$, $\usf{E}_2=\pr{0,1,0}$, and $\usf{E}_3=\pr{0,0,1}$. The blue cuboid marked $A_3$ denotes the accelerometer \#3. Its reference position vector is $\lsc{3}\usf{X}$. The components of $\lsc{3}\usf{X}$ w.r.t. $\pr{\usf{E}_i}_{i\in \mathcal{I}}$ are $\pr{-0.077, 0.016, -0.042}{\rm m}$. Similarly, the components of the other IMUs' reference position vectors are $\lsc{1}\usf{X}=\pr{0.077, 0.016, -0.042}{\rm m}$, $\lsc{2}\usf{X}=\pr{-0.001, -0.008, 0.015}{\rm m}$, and $\lsc{4}\usf{X}=\pr{0.001, 0.076, -0.078}{\rm m}$.  The arrow marked $\lsc{3}\usf{E}_i$, $i=1,2,3$, denote the IMU \#3's measurement directions in the reference configuration. Their components w.r.t. $\pr{\usf{E}_i}_{i\in \mathcal{I}}$ are $\pr{-0.446097, 0., -0.894985}$, $\pr{-0.241625, -0.962888, 0.120271}$, and $\pr{-0.86177, 0.269903, 0.429541}$. Similarly, the components of $\pr{\lsc{1}\usf{E}_i}_{i\in \mathcal{I}}$ are $\pr{\pr{0.444, 0., -0.895},\pr{-0.241, 0.962, -0.120},\pr{0.862, 0.269, 0.427}}$, of $\pr{\lsc{2}\usf{E}_i}_{i\in \mathcal{I}}$ are $\pr{\pr{0, -0.990, 0.134},\pr{1, 0, 0},\pr{0, 0.134, 0.990}}$, and of $\pr{\lsc{4}\usf{E}_i}_{i\in \mathcal{I}}$ are $\pr{\pr{0, 0.095, -0.995},\pr{-1, 0, 0},\pr{0, 0.995, 0.095}}$.
    Subfigures (c), (d), and (e) show the acceleration measurement components of time from sensors 1, 3, and 4, respectively. Subfigure (f) shows the angular velocity measurement components of time from sensor 1. $g=9.8~{\rm m/s^2}$. Each subfigure shows three graphs, labeled $i=1$, $i=2$, and $i=3$. These, respectively, correspond to the acceleration (or angular velocity) components measured by the accelerometer (or gyroscope) in its 1, 2, and 3, directions (see, e.g., the three arrows attached to the cuboid in the inset in (c)).
  }
    \label{fig:testov1}
\end{figure}

To recall, the A3G1-algorithm enables determination of the acceleration of a rigid body from measurements obtained by three accelerometers and one gyroscope.
In the present analysis, we use measurements from accelerometers \#1, \#3, and, \#4, together with those from gyroscope \#1.
Following Algorithm~\ref{algo:acce}, the inputs consist of discrete sequences of accelerations $\pr{\lsc{\ell}\alpha_i\ag{\cdot}}_{i \in \mathcal{I}}$ for $\ell\in \pr{1,3,4}$ and angular velocities $\pr{\lsc{1}\omega_i\ag{\cdot}}_{i \in \mathcal{I}}$ from the gyroscope.
The algorithm additionally requires the relative positions and orientations of the accelerometers, as well as the orientation of the gyroscope, $\lsc{\ell}\usf{X}$ and $\pr{\lsc{\ell}\usf{E_i}}_{i\in\mathcal{I}}$, $\ell\in \pr{1,3,4}$.
These parameters are provided in the caption of Fig.~\ref{fig:testov1}.

Upon supplying the sensor measurements and associated experimental parameters to Algorithm~\ref{algo:acce}, the A3G1-algorithm produces a discrete estimate of the acceleration at the point of interest.
As an illustration, in \S\ref{subsec:results} we present the predicted acceleration at the location of accelerometer \#2.

\subsection{Results}
\label{subsec:results}
\begin{figure}[h!]
    \centering
\includegraphics[width=0.9\textwidth]{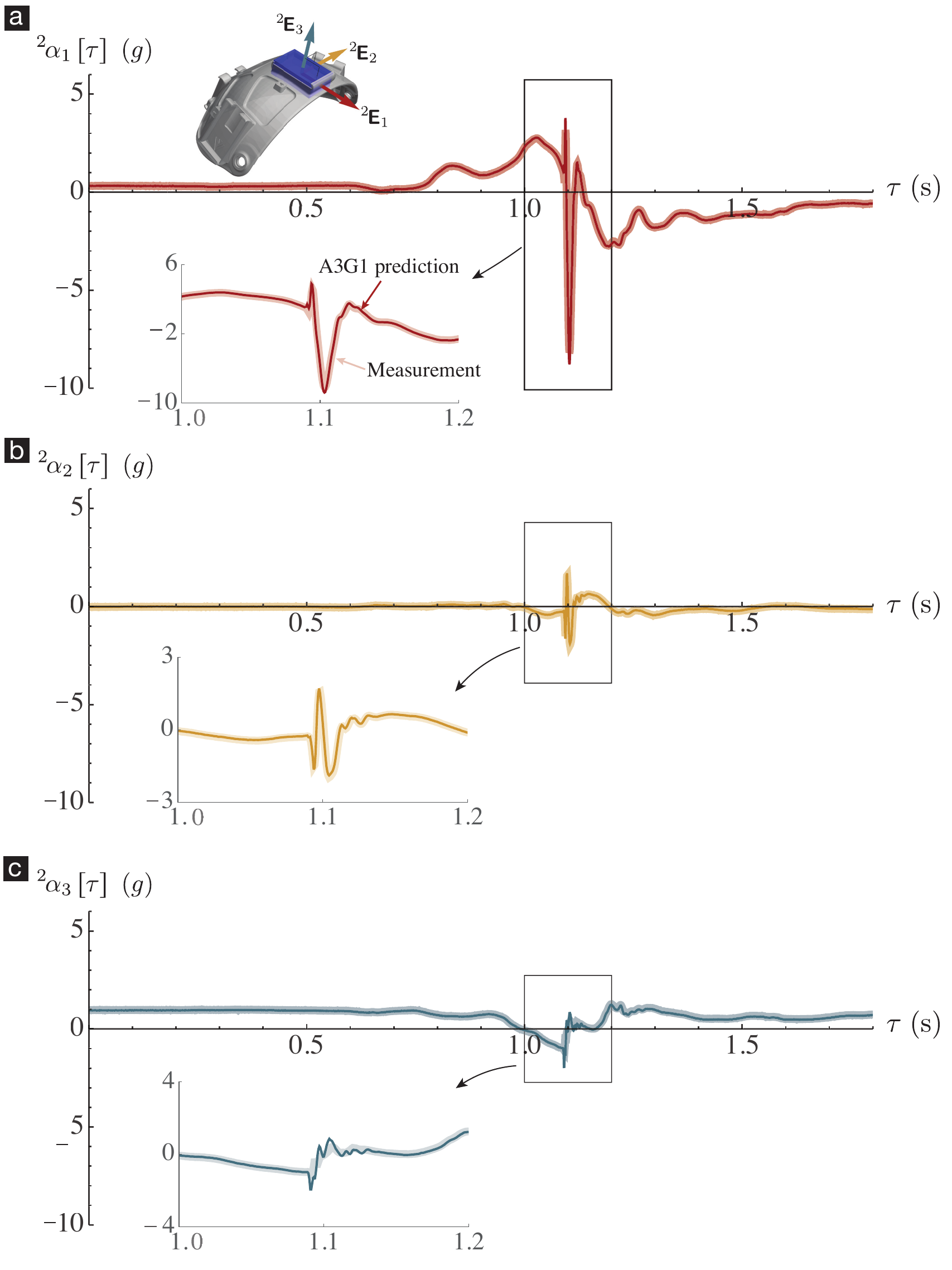}
    \caption{Comparison of predicted acceleration components of positions where we attached accelerometer 2 from A3G1 algorithm with the measurements from  accelerometer 2. $g=9.8~{\rm m/s^2}$}
    \label{fig:testpre}
\end{figure}

To validate the A3G1 algorithm, we predicted the acceleration of the point where we attached the sensor 2 (shown in Fig.~\ref{fig:testpre}(a)).
We then compare the predicted acceleration with the acceleration measurements from sensor 2.
The measurement axes of sensor 2, $\pr{\lsc{2}\usf{E}_i}_{i\in \mathcal{I}}$, is shown in the inset of Fig.~\ref{fig:testpre}(a).
we calculated the acceleration
by feeding the raw acceleration measurements from accelerometers \#1, \#3, and \#4, and the raw angular velocity measurements from gyroscope \#1.


The predicted acceleration components of sensor 2 in Test \#3, $\lsc{2}\alpha_i$ of time $\tau$, w.r.t. sensor 2's measurement axes $\lsc{2}\usf{E}_i$, $i \in \mathcal{I}$, are shown in red in Fig.~\ref{fig:testpre}(a), (b), and (c), respectively. The acceleration measurement from sensor 2 is shown in gray in Fig.~\ref{fig:testpre}.

We quantify the discrepancy between the measured and predicted acceleration using the error metric
\begin{equation}
\frac{\norm{\rm{P}\pr{\lsc{2}\busf{A}}\ag{\cdot}-\lsc{2}\busf{A}\ag{\cdot}}_{L^2(0,T)}}{\norm{\lsc{2}\busf{A}\ag{\cdot}}_{L^2(0,T)}},
\label{equ:error}
\end{equation}
where $\mathbb{R}\ni\tau\mapsto\rm{P}\pr{\lsc{2}{\busf{A}}}\ag{\tau}\in\mathcal{M}_{3,1}(\mathbb{R})$, is A3G1-algorithms' predictions for $\mathbb{R}\ni\tau\mapsto\lsc{2}\busf{A}\ag{\tau}\in\mathcal{M}_{3,1}(\mathbb{R})$.
The norm $\norm{\cdot}_{L^2(0,T)}$, where $T>0$, is the standard $L^2$ function norm,
\begin{equation}
\norm{\lsc{2}\busf{A}}_{L^2(0,T)}:=\pr{\int_{0}^{T}\lVert \lsc{2}\busf{A}\ag{\tau}\rVert^{2}_2\,d\tau}^{1/2},
\label{eq:L2norm}
\end{equation}
and $\ag{0,T}$ is $\lsc{2}\busf{A}\ag{\cdot}$'s domain.
For Test \#3, the error is $0.062$ (or $6.2\%$). The errors of all five tests are shown in Table~\ref{tb:Errors}.

\setlength{\tabcolsep}{10pt}
\renewcommand{\arraystretch}{2}
\begin {table}[H]
  \caption {Error measures of predicted acceleration of accelerometer \#2 for five tests}
\begin{center}
\scalebox{0.9}{%
\begin{tabular}{cccccc}
\hline
Tests
& Test \#1
 & Test \#2
 & Test \#3
  & Test \#4
& Test \#5 \\
\hline
 Error   & 6.4\%
 & 10.5\%
  & 6.2\%
 & 5.8\%
 &11.2\%\\
\hline
\end{tabular}}
\end{center}
\label{tb:Errors}
\end{table}

\section{Concluding remarks}
\label{sec:ConcludingRemarks}
\begin{enumerate}
    \item The spatial arrangement of the accelerometers $\pr{\lsc{\ell}\usf{X}}_{\ell\in\mathcal{I}}$ can be fairly general except for the restriction that all three accelerometers can not lie on the same line. This restriction follows from the assumption that $\Dbsf$ has full column rank. The gyroscope can be located anywhere on the rigid body.
\end{enumerate}

\section*{Acknowledgements}
The authors gratefully acknowledge support from the Panther Program, Tiger Program, and the Office of Naval Research (Dr. Timothy Bentley) under grants N000142112044 and N000142112054.
\section*{Declaration of Competing Interest}
The authors declare that they have no known competing financial interests or personal relationships that could have appeared to influence the work reported in this paper.

\bibliography{references.bib}

\end{document}

%% file: definitions.tex
\usepackage[hidelinks]{hyperref}

\newtheoremstyle{DefStyle}
  {10pt} 
  {5pt} 
  {} 
  {} 
  {\bfseries} 
  {.} 
  {10pt} 
  {} 

\theoremstyle{DefStyle}

\newcommand{\Mc}{\mathcal{M}}

\newcommand{\Absf}{\bm{\mathsf{A}}}
\newcommand{\Bbsf}{\bm{\mathsf{B}}}

\newcommand{\Dbsf}{\bm{\mathsf{D}}}
\newcommand{\Hbsf}{\bm{\mathsf{H}}}

\newcommand{\Mbsf}{\bm{\mathsf{M}}}

\newcommand{\Qbsf}{\bm{\mathsf{Q}}}

\DeclareMathOperator*{\argmin}{argmin}

\newcommand{\ag}[1]{\left[ #1 \right]}
\renewcommand{\u}[1]{\boldsymbol{#1}}

\newcommand{\usf}[1]{\u{\mathsf #1}}
\newcommand{\busf}[1]{\overline{\u{\mathsf #1}}}

\newcommand{\pr}[1]{\left( #1 \right)}
\newcommand{\set}[2]{\left\{#1\, \big|\, #2\right\}}
\newcommand{\norm}[1]{\left\lVert #1\right\rVert}

\newcommand{\lsc}[1]{{}^{#1}\!\,}

\usepackage{amsmath}

\usepackage{tikz}

\usepackage{mathabx}

\usepackage{algorithm2e}
\usepackage{algpseudocode}
\algrenewcommand\algorithmicrequire{\textbf{Input:}}
\algrenewcommand\algorithmicensure{\textbf{Output:}}

\usepackage{multirow}

%% file: main.bbl
\begin{thebibliography}{10}
\expandafter\ifx\csname url\endcsname\relax
  \def\url#1{\texttt{#1}}\fi
\expandafter\ifx\csname urlprefix\endcsname\relax\def\urlprefix{URL }\fi
\expandafter\ifx\csname href\endcsname\relax
  \def\href#1#2{#2} \def\path#1{#1}\fi

\bibitem{TBIdata}
{TBI Data | Concussion | Traumatic Brain Injury | CDC Injury Center},
  \url{https://www.cdc.gov/traumaticbraininjury/data/index.html}, accessed:
  2023-07-21.

\bibitem{Maas2022}
A.~I. Maas, D.~K. Menon, G.~T. Manley, M.~Abrams, C.~{\AA}kerlund, N.~Andelic,
  M.~Aries, T.~Bashford, M.~J. Bell, Y.~G. Bodien, et~al., Traumatic brain
  injury: progress and challenges in prevention, clinical care, and research,
  The Lancet Neurology 21~(11) (2022) 1004--1060.

\bibitem{Marshall257}
S.~Marshall, M.~Bayley, S.~McCullagh, D.~Velikonja, L.~Berrigan, Clinical
  practice guidelines for mild traumatic brain injury and persistent symptoms,
  Canadian Family Physician 58~(3) (2012) 257--267.

\bibitem{Dikmen2017}
S.~Dikmen, J.~Machamer, N.~Temkin, Mild traumatic brain injury: Longitudinal
  study of cognition, functional status, and post-traumatic symptoms, Journal
  of Neurotrauma 34~(8) (2017) 1524--1530.

\bibitem{YEATES2010}
K.~O. YEATES, Mild traumatic brain injury and postconcussive symptoms in
  children and adolescents, Journal of the International Neuropsychological
  Society 16~(6) (2010) 953–960.

\bibitem{masel2010traumatic}
B.~E. Masel, D.~S. DeWitt, Traumatic brain injury: a disease process, not an
  event, Journal of neurotrauma 27~(8) (2010) 1529--1540.

\bibitem{xiong2013}
Y.~Xiong, A.~Mahmood, M.~Chopp, Animal models of traumatic brain injury, Nature
  Reviews Neuroscience 14~(2) (2013) 128--142.

\bibitem{CERNAK2005}
I.~Cernak, Animal models of head trauma, NeuroRX 2~(3) (2005) 410--422.

\bibitem{GAETZ2004}
M.~Gaetz, The neurophysiology of brain injury, Clinical Neurophysiology 115~(1)
  (2004) 4--18.

\bibitem{BRAMLETT2007125}
H.~M. Bramlett, W.~D. Dietrich, Progressive damage after brain and spinal cord
  injury: pathomechanisms and treatment strategies, Vol. 161 of Progress in
  Brain Research, Elsevier, 2007, pp. 125--141.

\bibitem{gadd1966use}
C.~W. Gadd, Use of a weighted-impulse criterion for estimating injury hazard,
  Tech. rep., SAE Technical Paper (1966).

\bibitem{chou1974analytical}
C.~C. Chou, G.~W. Nyquist, Analytical studies of the head injury criterion
  {(HIC)}, SAE Transactions (1974) 398--410.

\bibitem{takhounts2013development}
E.~G. Takhounts, M.~J. Craig, K.~Moorhouse, J.~McFadden, V.~Hasija, Development
  of brain injury criteria ({B}r{IC}), Stapp Car Crash Journal 57 (2013) 243.

\bibitem{newman1986generalized}
J.~A. Newman, A generalized acceleration model for brain injury threshold
  ({GAMBIT}), in: Proceedings of the 1986 International IRCOBI Conference on
  the Biomechanics of Impact, 1986.

\bibitem{hajiaghamemar2021multi}
M.~Hajiaghamemar, S.~S. Margulies, Multi-scale white matter tract embedded
  brain finite element model predicts the location of traumatic diffuse axonal
  injury, Journal of Neurotrauma 38~(1) (2021) 144--157.

\bibitem{bar2016strain}
E.~Bar-Kochba, M.~T. Scimone, J.~B. Estrada, C.~Franck, Strain and
  rate-dependent neuronal injury in a {3D} in vitro compression model of
  traumatic brain injury, Scientific Reports 6 (2016) 30550.

\bibitem{RafaelTBImodel}
R.~D. González-Cruz, Y.~Wan, A.~Burgess, D.~Calvao, W.~Renken, F.~Vecchio,
  C.~Franck, H.~Kesari, D.~Hoffman-Kim, Cortical spheroids show
  strain-dependent cell viability loss and neurite disruption following
  sustained compression injury, Plos one 19~(8) (2024) 1--19.

\bibitem{wan2023mechanics}
Y.~Wan, R.~D. Gonz{\'a}lez-Cruz, D.~Hoffman-Kim, H.~Kesari, A mechanics theory
  for the exploration of a high-throughput, sterile 3{D} $\textit{in vitro}$
  traumatic brain injury model, Biomechanics and modeling in mechanobiology
  23~(4) (2024) 1179--1196.

\bibitem{madhukar2019finite}
A.~Madhukar, M.~Ostoja-Starzewski, Finite element methods in human head impact
  simulations: a review, Annals of Biomedical Engineering 47~(9) (2019)
  1832--1854.

\bibitem{Carlsen2021}
R.~W. Carlsen, A.~L. Fawzi, Y.~Wan, H.~Kesari, C.~Franck, A quantitative
  relationship between rotational head kinematics and brain tissue strain from
  a {2-D} parametric finite element analysis, Brain Multiphysics 2 (2021)
  100024.

\bibitem{Massouros2014}
P.~G. Massouros, P.~V. Bayly, G.~M. Genin, Strain localization in an
  oscillating maxwell viscoelastic cylinder, International Journal of Solids
  and Structures 51~(2) (2014) 305--313.

\bibitem{wan2023brain}
Y.~Wan, W.~Fang, R.~W. Carlsen, H.~Kesari, A finite rotation, small strain 2{D}
  elastic head model, with applications in mild traumatic brain injury, Journal
  of the Mechanics and Physics of Solids 179 (2023) 105362.

\bibitem{upadhyay2022data}
K.~Upadhyay, D.~G. Giovanis, A.~Alshareef, A.~K. Knutsen, C.~L. Johnson,
  A.~Carass, P.~V. Bayly, M.~D. Shields, K.~Ramesh, Data-driven uncertainty
  quantification in computational human head models, Computer Methods in
  Applied Mechanics and Engineering 398 (2022) 115108.

\bibitem{zhan2021rapid}
X.~Zhan, Y.~Liu, S.~J. Raymond, H.~V. Alizadeh, A.~G. Domel, O.~Gevaert, M.~M.
  Zeineh, G.~A. Grant, D.~B. Camarillo, Rapid estimation of entire brain strain
  using deep learning models, IEEE Transactions on Biomedical Engineering
  68~(11) (2021) 3424--3434.

\bibitem{wu2022real}
S.~Wu, W.~Zhao, S.~Ji, Real-time dynamic simulation for highly accurate
  spatiotemporal brain deformation from impact, Computer Methods in Applied
  Mechanics and Engineering 394 (2022) 114913.

\bibitem{Jeneel2025}
J.~Kachhadiya, J.~Romero, S.~Kou, Y.~Wan, H.~Kesari, R.~Szalkowski, J.~Andrews,
  Lightweight wearable headband with flexible hybrid electronics for
  head-kinematic monitoring and mild traumatic brain injury risk detection,
  IEEE Sensors Letters 9~(4) (2025) 1--4.

\bibitem{tripathi2025laboratory}
A.~Tripathi, Y.~Wan, S.~Malave, S.~Turcsanyi, A.~L. Fawzi, A.~Brooks,
  H.~Kesari, T.~Snedden, P.~Ferrazzano, C.~Franck, et~al., Laboratory
  evaluation of a wearable instrumented headband for rotational head kinematics
  measurement, Annals of Biomedical Engineering (2025) 1--18.

\bibitem{Buice2018}
J.~M. Buice, A.~O. Esquivel, C.~J. Andrecovich, Laboratory validation of a
  wearable sensor for the measurement of head acceleration in men's and women's
  lacrosse, Journal of Biomechanical Engineering 140.

\bibitem{abrams2024}
M.~Z. Abrams, J.~Venkatraman, D.~Sherman, M.~Ortiz-Paparoni, J.~R. Bercaw,
  R.~E. MacDonald, J.~Kait, E.~D. Dimbath, D.~Y. Pang, A.~Gray, et~al.,
  Biofidelity and limitations of instrumented mouthguard systems for assessment
  of rigid body head kinematics, Annals of Biomedical Engineering (2024)
  2872--2883.

\bibitem{wan2022determining}
Y.~Wan, A.~L. Fawzi, H.~Kesari, Determining rigid body motion from
  accelerometer data through the square-root of a negative semi-definite
  tensor, with applications in mild traumatic brain injury, Computer Methods in
  Applied Mechanics and Engineering 390 (2022) 114271.

\bibitem{Camarillo2013}
D.~B. Camarillo, P.~B. Shull, J.~E. Mattson, R.~Shultz, D.~Garza, An
  instrumented mouthguard for measuring linear and angular head impact
  kinematics in american football, Annals of Biomedical Engineering 41 (2013)
  1939--1949.

\bibitem{ovaska1998noise}
S.~J. Ovaska, S.~Valiviita, Angular acceleration measurement: A review, in:
  IMTC/98 Conference Proceedings. IEEE Instrumentation and Measurement
  Technology Conference. Where Instrumentation is Going (Cat. No. 98CH36222),
  Vol.~2, IEEE, 1998, pp. 875--880.

\bibitem{alonso2005noise}
F.~Alonso, J.~Castillo, P.~Pintado, Application of singular spectrum analysis
  to the smoothing of raw kinematic signals, Journal of Biomechanics 38~(5)
  (2005) 1085--1092.

\bibitem{rahaman2020accelerometer}
M.~M. Rahaman, W.~Fang, A.~L. Fawzi, Y.~Wan, H.~Kesari, An accelerometer-only
  algorithm for determining the acceleration field of a rigid body, with
  application in studying the mechanics of mild traumatic brain injury, Journal
  of the Mechanics and Physics of Solids 143 (2020) 104014.

\bibitem{cardou2008estimating}
P.~Cardou, J.~Angeles, Estimating the angular velocity of a rigid body moving
  in the plane from tangential and centripetal acceleration measurements,
  Multibody System Dynamics 19~(4) (2008) 383--406.

\bibitem{cardou2009linear}
P.~Cardou, J.~Angeles, Linear estimation of the rigid-body acceleration field
  from point-acceleration measurements, Journal of Dynamic Systems,
  Measurement, and Control 131~(4) (2009) 041013.

\bibitem{padgaonkar1975measurement}
A.~J. Padgaonkar, K.~W. Krieger, A.~I. King, {Measurement of angular
  acceleration of a rigid body using linear accelerometers}, Journal of Applied
  Mechanics 42~(3) (1975) 552--556.

\bibitem{zou2018algorithm}
T.~Zou, J.~Angeles, An algorithm for rigid-body angular velocity and attitude
  estimation based on isotropic accelerometer strapdowns, Journal of Applied
  Mechanics 85~(6) (2018) 061010.

\bibitem{Xiaobo2025}
X.~Liu, A novel approach for computing rigid body motion using linear
  accelerations, Journal of Applied Mechanics 92~(9) (2025) 091003.

\bibitem{naunheim2003linear}
R.~Naunheim, P.~Bayly, J.~Standeven, J.~Neubauer, L.~Lewis, G.~Genin, Linear
  and angular head accelerations during heading of a soccer ball, Medicine \&
  Science in Sports \& Exercise 35~(8) (2003) 1406--1412.

\bibitem{golub2013matrix}
G.~H. Golub, C.~F. Van~Loan, Matrix computations, JHU press, 2013.

\end{thebibliography}
